\documentclass[structabstract]{aa}  
\usepackage{txfonts, natbib}
\usepackage{graphicx, amssymb}
\begin{document}
\noindent

\title{A search for line intensity enhancements in the far-UV spectra of active late-type stars arising from opacity}
\vskip1.0truecm

\author{F. P. Keenan\inst{1}, D. J. Christian\inst{2}, S. J. Rose\inst{3}, and M. Mathioudakis\inst{1} 
}

 \institute{Astrophysics Research Centre, School of Mathematics and Physics, Queen's University Belfast, Belfast BT7 1NN, Northern Ireland, UK
\\
\email{F.Keenan@qub.ac.uk}
\and
 Department of Physics and Astronomy, California State University, 
18111 Nordhoff Street, Northridge, CA
91330, USA
\and
Department of Physics, Imperial College, London SW7 2BZ, UK
} 

\abstract
   {
   Radiative transfer calculations have predicted intensity enhancements for optically thick emission lines, as opposed to the normal intensity reductions, for astrophysical plasmas under certain conditions. In particular, the results are predicted to be dependent both on the geometry of the emitting plasma and the orientation of the observer. Hence in principle the detection of intensity enhancement may provide a way of determining the geometry of an unresolved astronomical source.}
   {To investigate such enhancements we have analysed a sample of active late-type stars observed in the far ultraviolet spectral region.}
   {Emission lines of \ion{O}{vi} in the FUSE satellite spectra of $\epsilon$\,Eri, II\,Peg and Prox Cen were searched for intensity enhancements due to opacity. }
   {We have found strong evidence for line intensity enhancements due to opacity during active or flare-like activity for all three stars.  The \ion{O}{vi} 1032/1038 line intensity ratios, predicted to have a value of 2.0 in the optically thin case, are found to be up to $\sim$\,30\%\ larger during several orbital phases.  }
   {Our measurements, combined with radiative transfer models, allow us to constrain both the geometry of 
   the \ion{O}{vi} emitting regions in our stellar sources and the orientation of the observer. A spherical emitting plasma can be ruled out, as this would lead to no intensity enhancement. In addition, the theory tells us that the line-of-sight to the plasma must be close to perpendicular to its surface, as observations at small angles to the surface lead to either no intensity enhancement or the usual line intensity decrease over the optically thin value. For the future, we outline a laboratory experiment, that could be undertaken with current facilities, which would provide an unequivocal test of predictions of line intensity  enhancement due to opacity, in particular the dependence on plasma geometry. 
 }

\keywords{stars: activity --- stars: late-type --- opacity}

\authorrunning{F. P. Keenan et al.}
\titlerunning{Intensity enhancement due to opacity}

\maketitle

\section{Introduction}

Deriving the physical extent and geometry of spatially unresolved stellar corona and transition regions (TRs) is an important undertaking. Previous studies seeking to derive scale heights for stellar TRs 
(Mathioudakis et al. 1999; Bloomfield et al. 2002; Christian et al. 2004) used the relation of optical depth 
to physical parameters, including electron density and pathlength. Transitions such as 1032\,\AA\ (2s $^{2}$S--2p $^{2}$P$_{3/2}$) and 1038\,\AA\ (2s $^{2}$S--2p $^{2}$P$_{1/2}$)
of \ion{O}{vi}  
have a common lower level, and under optically thin conditions in a collisionally-excited plasma their 
expected line intensity ratio scales as that of their 
electron impact excitation rates, which is 2.0 (Aggarwal \& Keenan 2004). Hence deviations of the observed line ratio from the optically thin value
allows the optical depth and hence parameters of the emitting plasma to be derived.
Using this method, several studies have found TR scale heights between 10--100 km for active stars including  
AU\,Mic
(Bloomfield et al. 2002), the flare star Prox Cen (Christian et al. 2004) and AD\,Leo (Christian et al. 2006). 

 We note that 1032/1038 intensity ratios of \ion{O}{vi} with values ranging from 1--4 have been measured in
solar spectra obtained by the Ultraviolet Coronagraph Spectrometer (UVCS) on the SOHO satellite (see, for example, Nakagawa 
2008). However, these observations were obtained at large heights above the solar surface, where the electron densities
are very low (around 10$^{7}$\,cm$^{-3}$ or less; Ko et al. 2006), and the \ion{O}{vi} line emission has a significant component from the resonant scattering of chromospheric \ion{O}{vi} radiation and/or the absorption and subsequent re-emission of Doppler-shifted \ion{C}{ii} 1036.3 and 1037.0\,\AA\ photons by \ion{O}{vi} 1038\,\AA\ (Kohl \& Withbroe 1982). 
The \ion{O}{vi} line intensities decrease rapidly with increasing height above the solar surface, with for example a factor of 10 reduction between 1 and 1.1\,R$_{\odot}$ (Nagakawa 2008). Hence for an unresolved stellar source, such as those considered here, the \ion{O}{vi} line emission will be dominated by the high density collisionally-excited component and will have an optically thin 1032/1038 intensity ratio of 2.0. 

Recently, theoretical work has indicated that in certain instances the intensity ratio of an optically thick to optically thin
line in a collisonally-excited plasma could increase (rather than the expected decrease), due to the fact that an ion in the upper state of the transition can be pumped in the optically thick case by photons traversing the plasma at many different angles 
(Kerr et al. 2004, 2005). 
 In addition, these studies predicted that the degree of line
enhancement for the optically thick transition depends on both the geometry of the
emitting plasma and the orientation of the observer. This is an important result, as it potentially provides
a way of determining the geometry of a spatially unresolved plasma using purely spectroscopic means.
Some evidence for such a line enhancement was recently found for the active star EV\,Lac 
using observations of the 15.01 and 16.78\,\AA\ lines of \ion{Fe}{xvii} from the XMM-Newton satellite. 
Rose et al. (2008) measured a 15.01/16.78 intensity ratio of 2.50 $\pm$ 0.25 (1$\sigma$ error), compared to a theoretical optically thin value of 
between 1.75--1.93, indicating some enhancement in the optically thick 15.01\,\AA\ transition. However, we should note that there are uncertainties in the atomic data for \ion{Fe}{xvii} and the theoretical line ratios are also sensitive to the adopted plasma conditions (see Gillaspy et al. 2011 and references therein).

In the present work, we extend our study to far-ultraviolet observations of active late-type stars.
Specifically, we use moderate  to high resolution spectra from the Far Ultraviolet Spectroscopic
Explorer (FUSE) satellite to search for line intensity enhancements due to opacity in the atmospheres of several late-type stars.  We examine the 1032 and 1038\,\AA\ emission lines of \ion{O}{vi}, where the transitions have a common lower level and the theoretical value of the optically thin 1032/1038 ratio involves simpler and more reliable atomic physics, plus has a lesser dependence on the plasma parameters, than is the case for \ion{Fe}{xvii}. 
In Section 2 we describe the FUSE observations and data reduction techniques, while Section 3 provides details of the theory used in the analysis.  
Our results are presented in Section 4, including measurements of the 
emission line fluxes, and the determination of densities and pathlengths for the emitting plasmas under consideration. In Section 5 
we discuss our line ratios and implications for the geometry of the emitting region, and also 
outline some possibilities for future work, including a laboratory experiment to test the theory 
behind this research.

\section{Observations}

The FUSE satellite operated successfully from its launch in June 1999 until October 2007.  
It consisted of four co--aligned prime--focus telescopes, with
two having SiC coatings and optimised for the 905--1105\,\AA\ region, 
and the others using LiF coatings to cover 987--1187\,\AA. Holographically-ruled 
gratings were used to disperse light onto two separate microchannel                            
plate detectors (MCPs), each with two independent segments. The spectral resolution for the HIRS 
aperture was $\sim$\,20,000 and 
$\sim$\,12,000--15,000 for LWRS. Details of the FUSE instrument and    
in-orbit performance may be found in Moos et al. (2000) and Sahnow et al. (2000a,b).  

\subsection{Sample selection} 

The FUSE observations and datasets were obtained from the Multimission
Archive at Space Telescope (MAST).  Quick-look spectra for the \ion{O}{vi} lines
(1032 and 1038\,\AA) 
were investigated and stars with possible line ratio enhancements selected. 
These include the active star $\epsilon$\,Eri and the RS CVn II\,Peg. In addition, we employ \ion{O}{vi} line ratio results for Prox Cen from Christian et al. (2004). Table 1 summarises the observational 
datasets for the stellar sample used in the current study.

\begin{center}
\begin{table*}
\caption{Summary of observations for the stellar sample.}
\begin{tabular}{lcccccr}
\hline
Source     & Program ID  & Exp$^a$    &  Date$^b$  &  aperture & Spectral type  & Comment  \\
\hline
Prox Cen    & D122     & 45.4   &  2003-04-05    &  LWRS     &  M5.5 Ve    & Flare star   \\ 
 $\epsilon$ Eri       & C165          &  73      & 2003-12-21        &  HIRS     &       K2 V   &   \\
II Peg   & P179  &  58       &  2000-11-23       &  LWRS     &  K0 IV          &  RS CVn  \\
\hline
\end{tabular}
\label{tab1}
\medskip
\\
$^{a}$Total exposure time in ksec.
\\
$^{b}$Observation date (earliest date given if more than 1 sequence).

\end{table*}

\end{center}

\subsection{Data analysis} 

The raw FUSE data files were re-processed with the latest version
of the FUSE calibration pipeline (currently CalFUSE V3.2).
Spectra were extracted from the appropriate aperture (HIRS or LWRS)
and background subtracted, flat-fielded, wavelength and flux calibrated 
in the standard CalFUSE reduction (Dixon et al. 2007).
The $\epsilon$\,Eri spectra were obtained in the high resolution mode (HIRS), 
and II\,Peg in the LWRS mode. 

Spectra were extracted as a function of intensity and phase using the new IDF\_CUT routine.  The first 25 ksec  of the $\epsilon$\,Eri observations were a factor of 25 brighter than the remaining ones (see Fig. 1), which had a median count rate of 0.009 counts/sec. This initial 25 ksec was divided into 8 phase bins. Most of the II\,Peg observations
were active and this entire dataset was divided into 11 phase bins. 
Extracted spectra were fitted using IRAF and DIPSO routines. 

As in our previous work (Bloomfield et al. 2002; Christian et al. 2004, 2006) we measured emission
line fluxes by fitting Gaussians to the observed line profiles. 
In general, a single Gaussian profile was sufficient to fit both the \ion{O}{vi} 
transitions. 

Light curves extracted in the lines 
of \ion{O}{vi} (1032 + 1038\,\AA)
are shown in Figs. 1 and 2 for  $\epsilon$\,Eri and II\,Peg, respectively.

\begin{figure*}
\includegraphics[scale=0.45,angle=0]{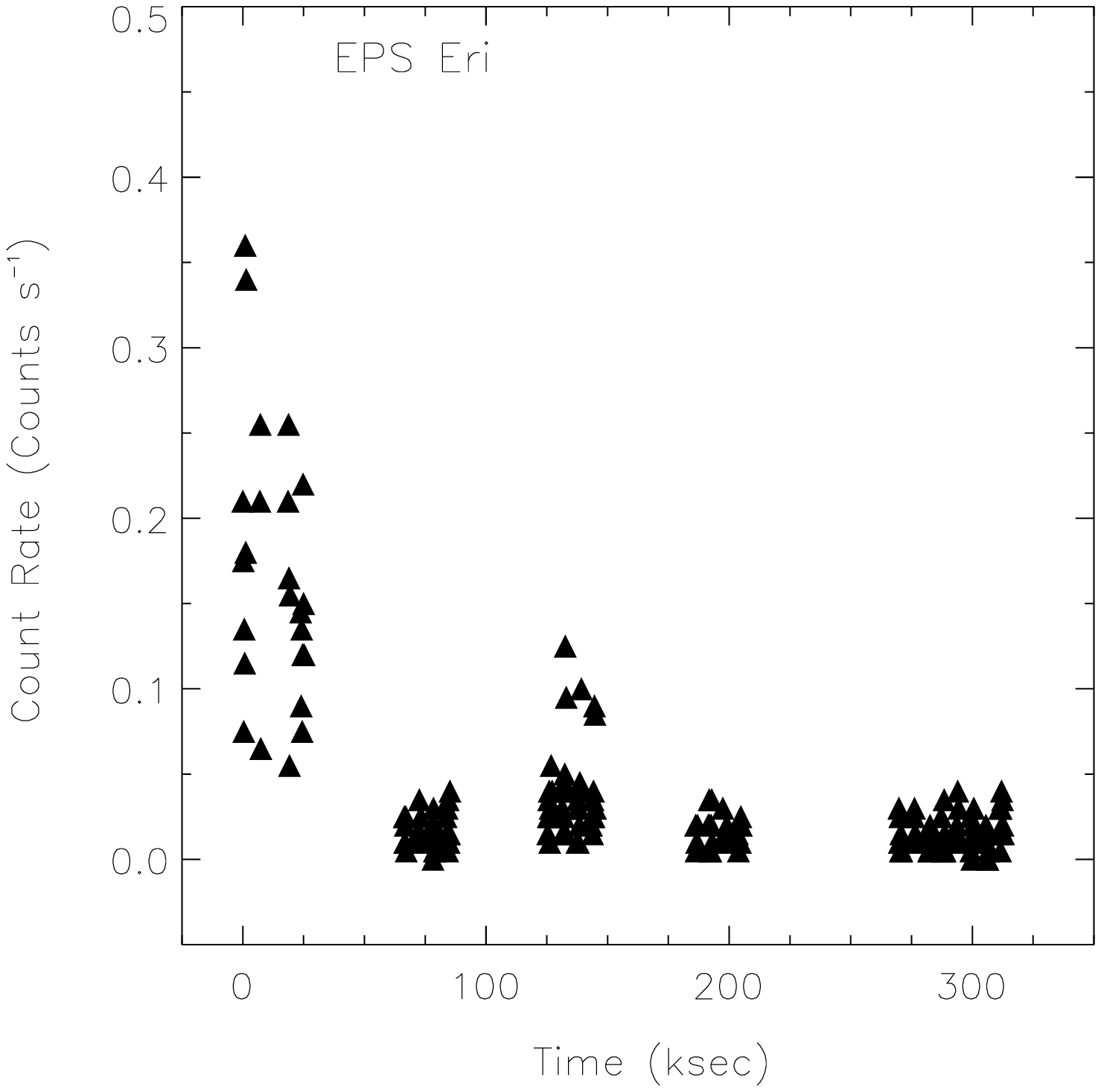}
\includegraphics[scale=0.45,angle=0]{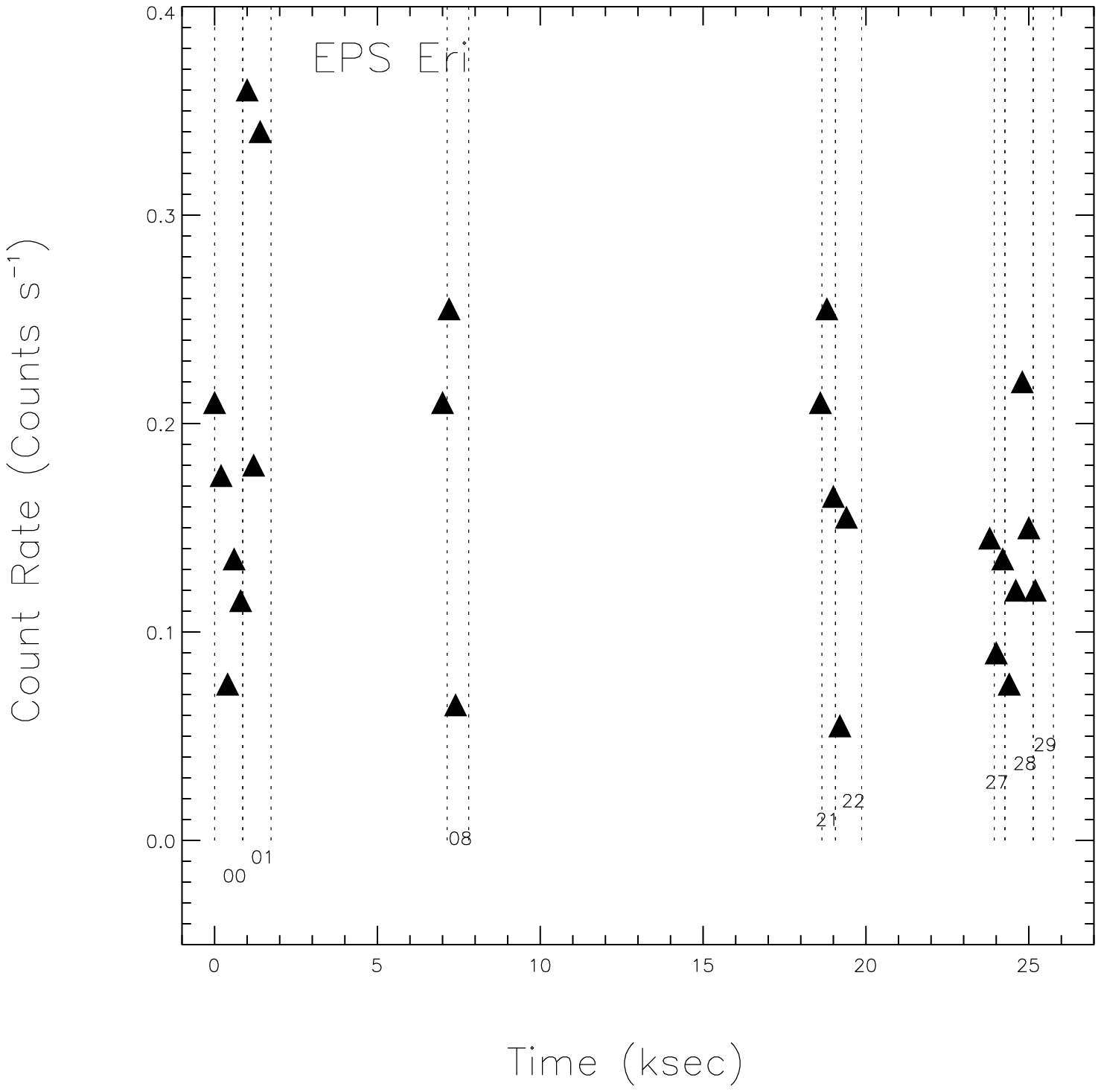}
\caption{Light curves for $\epsilon$ Eri over the entire time period of the FUSE observations.
The left-hand panel show the \ion{O}{vi} light curve (1032 + 1038\,\AA\ lines) and the right-hand one
an 
expanded view of the first 25 ksec, which has the greatest level of stellar activity.
Labels for phases 00 to 29, for intervals that had non-zero exposure of good signal-to-noise, are given and correspond to the Specnum names in the last column of Table 2; i.e. phase 02 did not have sufficient signal and is not in Table 2.  The bin size is 200 sec for both light curves. Spectral bins are indicated by the
dashed vertical lines.
}
\end{figure*}

\begin{figure*}
\includegraphics[scale=0.45,angle=0]{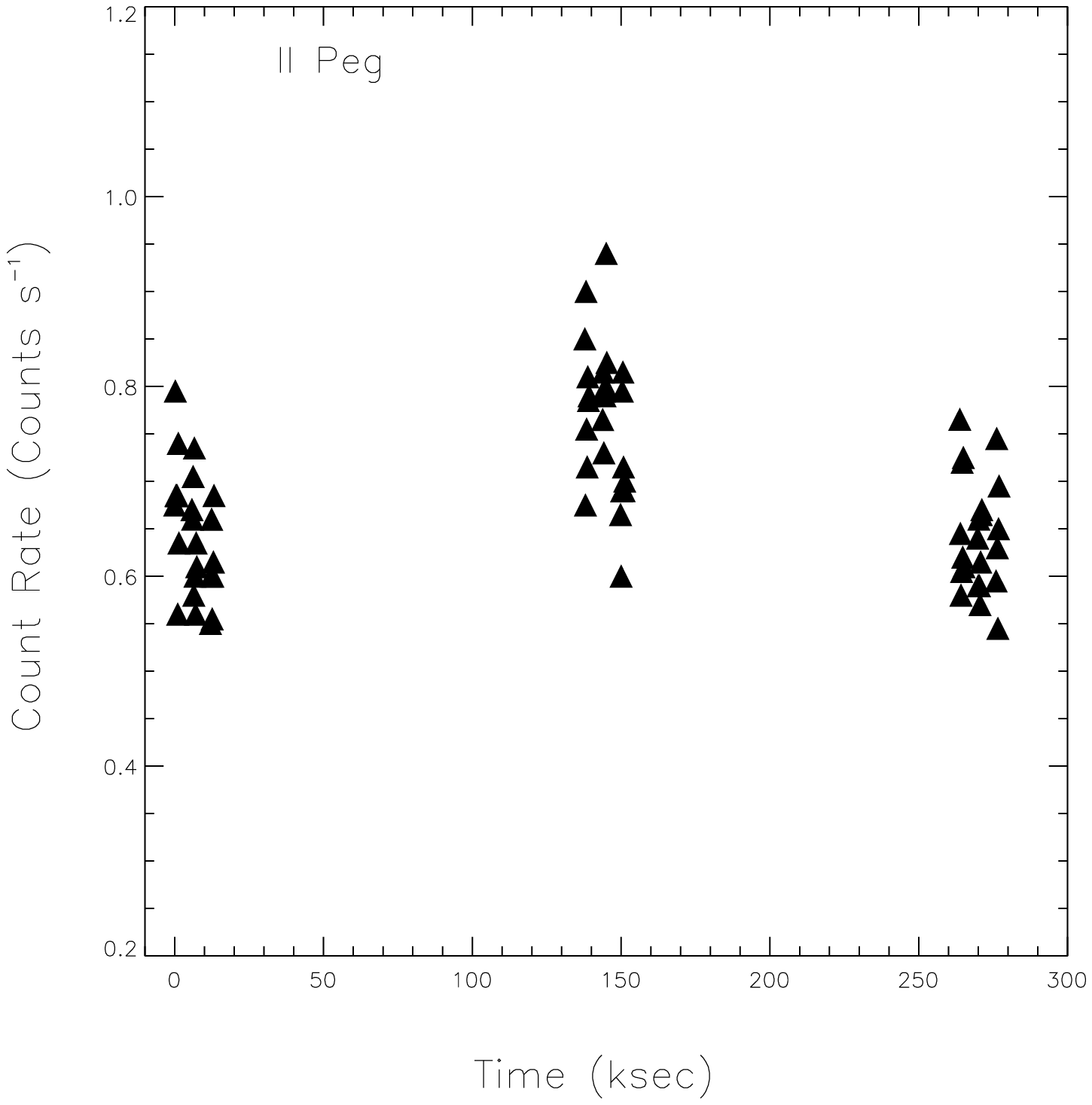}
\includegraphics[scale=0.45,angle=0]{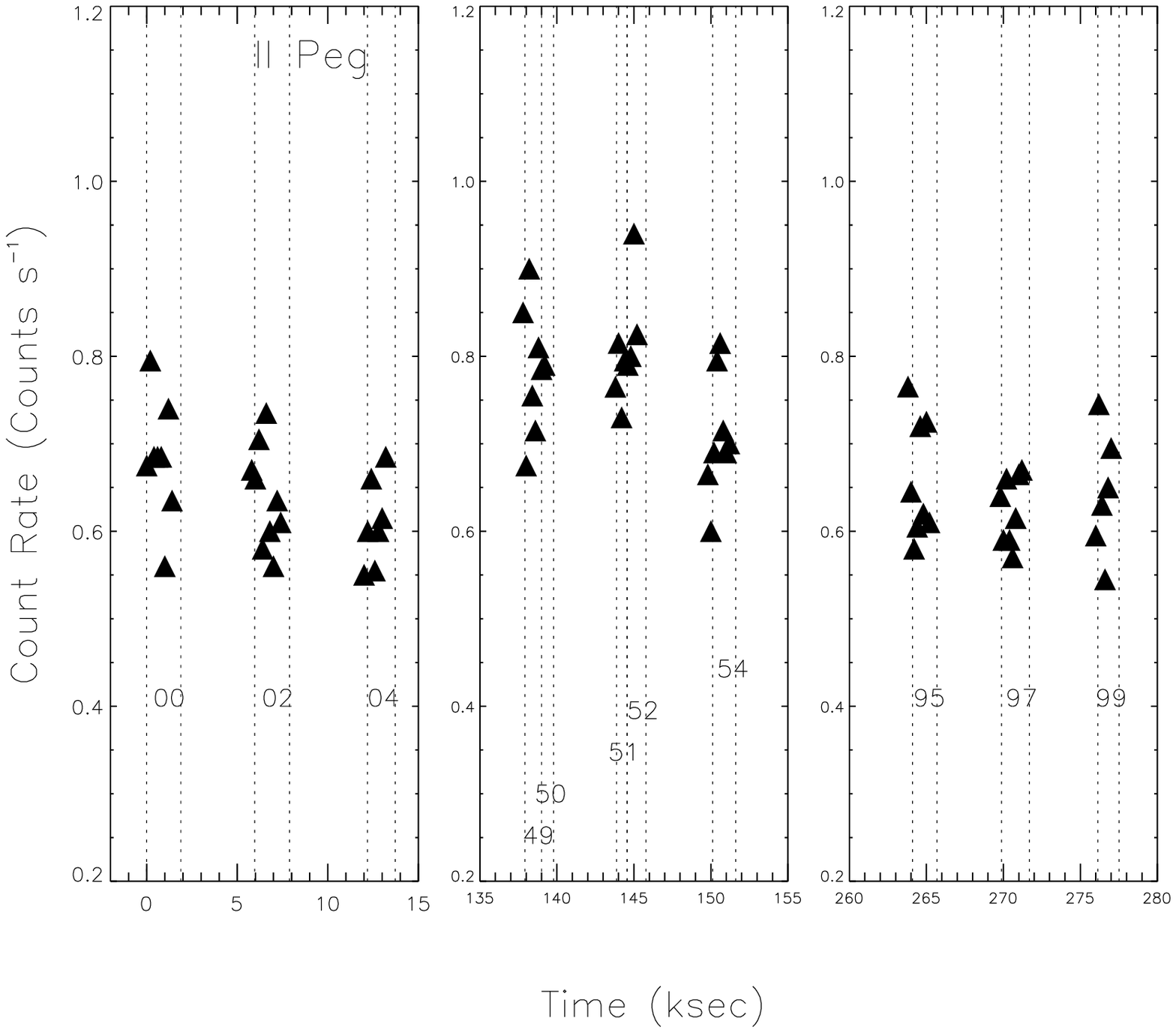}
\caption{Light curves in the \ion{O}{vi} lines (1032 + 1038\,\AA) for II\,Peg. 
The left-hand panel shows the entire time period of the FUSE observations, while the right shows an
expanded portion of the
individual observations, plus vertical dashed lines for the exposure windows from IDF\_CUT.
Labels for phases p00 to p99 are given and correspond to the Specnum names in Table 2.
The bin size is 200 sec for both light curves.
}
\end{figure*}

\section{Theory}

In this section we extend the previous theoretical work by Kerr et al. (2005) on the variation of 
line intensity enhancement with optical depth, and its relation to the 
plasma geometry. The research of Kerr et al. made several assumptions, including uniform plasma conditions and specific geometries. However, the lines they considered involved different kinetic pathways to the two transitions
discussed here and so we need to alter their analysis in this respect. Specifically, Kerr et al. investigated
one optically thick and one optically thin line which did not involve any common states. In
our case, the two lines of Li-like \ion{O}{vi} at 1032 and 1038\,\AA\ 
involve a common lower level (the ground state 2s $^{2}$S), 
with oscillator strengths and hence optical depths 
differing only by a factor of two. The densities and temperatures involved are such that the kinetic system can be considered to be in coronal steady-state, with the upper two levels (2p $^{2}$P$_{1/2}$ and $^{2}$P$_{3/2}$) not connected kinetically. If an analysis similar to that undertaken by Kerr et al. is applied to this problem, and noting that the electron collisional excitation rates for the two transitions 
2s $^{2}$S--2p $^{2}$P$_{3/2}$ (1032\,\AA) and 
2s $^{2}$S--2p $^{2}$P$_{1/2}$ (1038\,\AA) are predicted to differ by a factor of two (Aggarwal 
\&\ Keenan 2004), then the line intensity ratio is:

\[\frac{I(1032)}{I(1038)} = 2 \frac{g_a (1032) g_b (1038)}{g_b (1032) g_a (1038)}\]

where $g_a (k)$  is the probability for photon escape in the line-of-sight of the observer for line $k$ and 
$g_b (k)$ is the angle-averaged escape probability. 
 Both angle specific and angle-averaged escape probabilities (also known as escape factors) are required, with the former determining the transport of photons to the observer, which 
depends on the specific angle of observation. The angle-averaged escape factor
controls the pumping of the upper state by photons from the rest of the plasma, which in turn 
determines the number of excited states that can emit a photon, and this requires an average over all angles. It is the inclusion of both these factors that is needed to understand why the line ratio can exceed the optically thin limit. 

At first sight one might think that if the upper state is pumped by the photon field, this involves the destruction of the photon doing the pumping and would appear to suggest that there is no real gain in the population of the upper level. However, literature of the escape factor method of accounting for photon absorption in a line or lines shows that this is not the case (see for example Mihalas 1970). Consequently, we maintain that our use of 
an escape factor approach coupled to a collisional radiative model takes into account all the relevant physics and is entirely realistic. This includes collisional excitation (although de-excitation is much slower than spontaneous radiative de-excitation as we are in collisional-radiative steady-state) and photo-excitation. Indeed, the latter is the process by which photons are absorbed by the plasma and that is accounted for by the escape factor method.

For the case of a spherical plasma, $g_a (k)$ = $g_b (k)$ by symmetry, and hence I(1032)/I(1038) = 2. However, for an infinite plane slab the factor becomes:

\[\frac{I(1032)}{I(1038)} =\]

\[2 \frac{\int_{0}^{\infty} (1 - exp [- \frac{2 \tau_0}{\mu^{\prime}} e^{x^{2}} ]) dx \int_{0}^{\infty} \int_{0}^{1} \mu (1 - exp [- \frac{\tau_0}{\mu} e^{x^{2}} ]) d \mu dx}{\int_{0}^{\infty} (1 - exp [- \frac{\tau_0}{\mu^{\prime}} e^{x^{2}} ]) dx \int_{0}^{\infty} \int_{0}^{1} \mu (1 - exp [- \frac{2 \tau_0}{\mu} e^{x^{2}} ]) d \mu dx}\]

where $\tau_0$ is the optical depth of the 1038\,\AA\ transition for  a pathlength ($l$) which is the perpendicular thickness of the plasma slab, and $\mu^{\prime}$ = cos $\theta$ where $\theta$ is the angle between the perpendicular to the slab and the line-of-sight to the observer (see Fig. 3).  This expression is evaluated by numerical integration, and our results are shown in Fig. 4 and discussed in Section 5. However, in Fig. 4 we plot the intensity ratio as a function of 
column density $n_e l$ (i.e. product of electron density and pathlength) rather than $\tau_0$ as the former is 
a commonly-employed quantity and is relatively straightforward to at least approximately estimate (see Section 4.4).  

Although our results formally only hold for an infinite slab, in reality they will be applicable to any case where the lateral dimension of the emitting plasma is much greater than the thickness {\em l}. Our theoretical models should therefore be a good approximation for the stellar transition regions considered here, given the small values of {\em l} found for these (see Sections 1 and 4.4), especially compared to the radius of a star.

As the purpose of the current work is to 
illustrate the principle of this type of analysis we do not calculate the line intensity ratio for other geometries. However, we note that this will be the subject of a later paper.

\begin{figure}
\includegraphics[scale=0.8,angle=0]{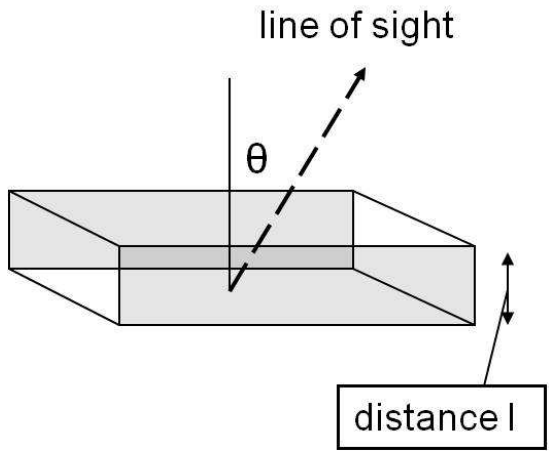}
\caption{Schematic diagram of the infinite plane slab geometry discussed in Section 3. The line-of-sight to the observer is at an angle $\theta$ to the perpendicular to the slab, which is of thickness $l$. 
}
\end{figure}

\begin{figure*}
\includegraphics[scale=0.45,angle=0]{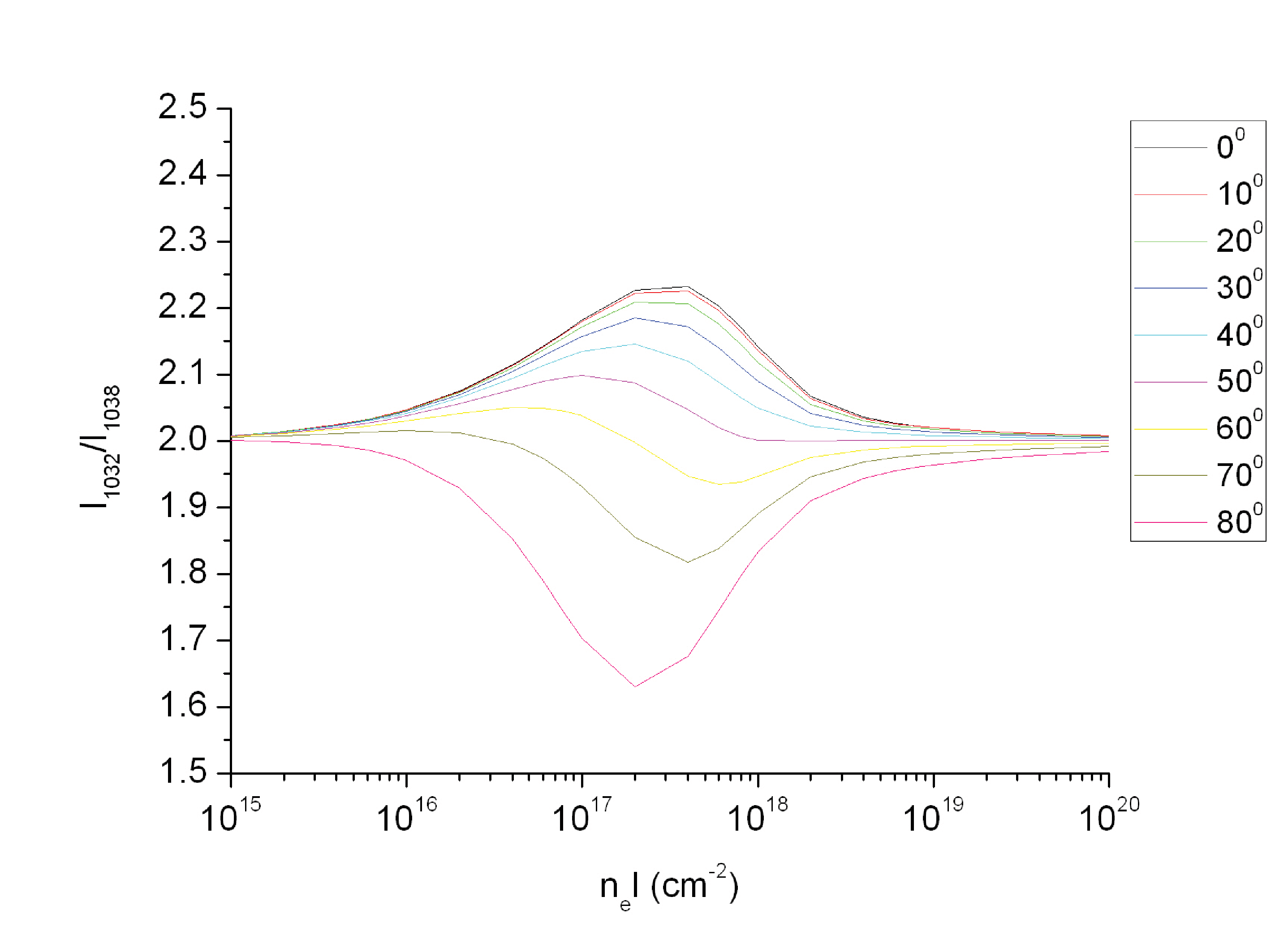}
\caption{Plot of the \ion{O}{vi} 1032/1038 emission line intensity ratio as a function of column density $n_e l$ for the infinite plane slab plasma illustrated in Fig. 3, for various values of angle of observation $\theta$. 
}
\end{figure*}

\section{Results}

 We have searched for line intensity enhancements in several active late-type stars observed with FUSE, employing the
 the \ion{O}{vi} 1032/1038  
 line ratio as a diagnostic. Measured \ion{O}{vi} line fluxes and 1032/1038 intensity ratios,  with their associated 1$\sigma$ errors, are listed in Table 2.   
  Below we discuss the measurements for each star and their variation as a function of orbital phase.

\begin{center}
\begin{table*}
\caption{FUSE \ion{O}{vi} 1032 and 1038\,\AA\ fluxes and emission line intensity ratios.}
\begin{tabular}{llccccccr}
\hline
Phase &  \multicolumn{2}{c}{Time}   &   \multicolumn{2}{c} {\ion{O}{vi} flux}    & Ratio & Comment    \\
           &       MJD & T$_{exp}$             & (1032 \AA) & (1038 \AA) & & \\
  &  (day) & (sec) &  10$^{-13}$ erg cm$^{-2}$ s$^{-1}$ &  10$^{-13}$ erg cm$^{-2}$ s$^{-1}$ & 1032/1038 & Specnum \\
\hline

{\bf $\epsilon$ Eri}
\\
\ldots & 52993.4890   &  51816 &  0.160$\pm$0.003 & 0.080$\pm$0.003 &  1.93$\pm$0.08 & Quiet
\\                 
 0.903 & 52992.7305  & 867  & 0.91$\pm$0.03 & 0.54$\pm$0.03  & 1.67$\pm$0.10  & 00 
 \\
 0.903 & 52992.7383  & 866 & 1.44$\pm$0.03 & 0.64$\pm$0.04  & 2.26$\pm$0.16  & 01  
 \\
 0.909 &  52992.8086 &  659 &  1.06$\pm$0.04 & 0.41$\pm$0.03 & 2.58$\pm$0.20 & 08 
 \\
 0.921 &  52992.9375 & 418 & 1.02$\pm$0.04 & 0.56$\pm$0.04  & 1.81$\pm$0.15 & 21 
  \\
 0.922 &  52992.9492  & 801 & 0.99$\pm$0.04 & 0.60$\pm$0.04 & 1.67$\pm$0.13 & 22 
 \\
 0.926 &  52993.0001     & 431 & 0.67$\pm$0.03 & 0.25$\pm$0.05  & 2.72$\pm$0.54 & 27, low S/N 
 \\
 0.927 &  52993.0117    & 866 & 0.72$\pm$0.03 & 0.37$\pm$0.03  & 1.95$\pm$0.20 & 28 
 \\
 0.928 &   52993.0195    & 626 & 1.09$\pm$0.03 & 0.48$\pm$0.03  & 2.27$\pm$0.15 &  29 
 \\
  \\
{\bf II Peg } 
\\ 
\ldots & 51871.670   & 15739 &   3.03$\pm$0.04 & 1.49$\pm$0.04 & 2.03$\pm$0.07 & Total
\\
0.443 & 51871.681    & 1886 & 2.92$\pm$0.08 & 1.36$\pm$0.08 & 2.14$\pm$0.14  & 00 
\\
0.454 &51871.750     & 1919 &  2.70$\pm$0.07 & 1.22$\pm$0.07  & 2.21$\pm$0.13 & 02 
\\
0.464 & 51871.820    & 1532 & 2.68$\pm$0.05 & 1.16$\pm$0.05  & 2.30$\pm$0.11  & 04 
\\
0.680 & 51873.272    & 1074 & 3.54$\pm$0.13 & 1.68$\pm$0.18  & 2.11$\pm$0.24  & 49
 \\
0.682 & 51873.283  &  786  & 3.40$\pm$0.11  & 1.42$\pm$0.12 & 2.40$\pm$0.21 & 50  
\\
0.690 & 51873.339  &  687  & 3.19$\pm$0.11  & 1.54$\pm$0.12 & 2.07$\pm$0.17  & 51 
\\
0.692 & 51873.350  & 1221 & 3.48$\pm$0.11  & 1.82$\pm$0.11 & 1.91$\pm$0.13  & 52  
\\
0.701 & 51873.416  & 1496 & 3.15$\pm$0.08  & 1.50$\pm$0.09 & 2.10$\pm$0.14  & 54 
\\
0.897 & 51874.735  & 1585 & 2.83$\pm$0.08  & 1.41$\pm$0.08 & 2.00$\pm$0.12  & 95 
\\
0.908 & 51874.804  & 1805 & 2.70$\pm$0.07  & 1.28$\pm$0.08 & 2.11$\pm$0.14  & 97 
\\
0.918 & 51874.874  & 1380 & 2.57$\pm$0.09  & 1.48$\pm$0.10 & 1.74$\pm$0.14  & 99 
\\
\\
{\bf Prox Cen} 
\\
Quiet & 52734.4748 & 39840 & 1.21$\pm$0.03 & 0.66$\pm$0.03 &1.85$\pm$0.11 & Christian et al. (2004)
\\
Flare 1 & 52734.4748 & 600 & 5.50$\pm$0.15 & 2.44$\pm$0.15 & 2.25$\pm$0.15 & Christian et al. (2004) 
\\
Flare 2 & 52735.1594 &1200 & 0.97$\pm$0.05  & 0.41$\pm$0.06  & 2.40$\pm$0.35 & Christian et al. (2004)
\\
\hline
\end{tabular}
\end{table*}
\end{center}

 \subsection{$\epsilon$ Eri}
 
 FUSE spectra for $\epsilon$\,Eri extracted from the initial part of the observations showed 1032/1038 intensity ratios  greater than the optically thin value 
(2.0) at statistically significant levels. For example,
we find a measured line ratio of 2.58 $\pm$ 0.20 for Specnum 08 (see Table 2), which is larger than the theoretical value by 2.9$\sigma$, indicating a 99.6\%\ confidence level in our result being a true enhancement.
We computed the orbital phase of each spectrum using the ephemeris of $\epsilon$\,Eri and rotational period of 11.35 days from Croll et al. (2006), and the FUSE observations covered orbital phases from 0.90--0.93.
The enhancement was observed for phases 0.903 and 0.93, and marginally for phase 0.909. However, spectra from intervals at later times (after 25 ksec) revealed no 1032/1038 ratios greater than 2.0. 
We show sample \ion{O}{vi} spectra for $\epsilon$\,Eri in Fig. 5, while the  1032/1038 ratios 
are plotted as a function of orbital phase in Fig. 6.

\begin{figure*}
\includegraphics[scale=0.45,angle=0]{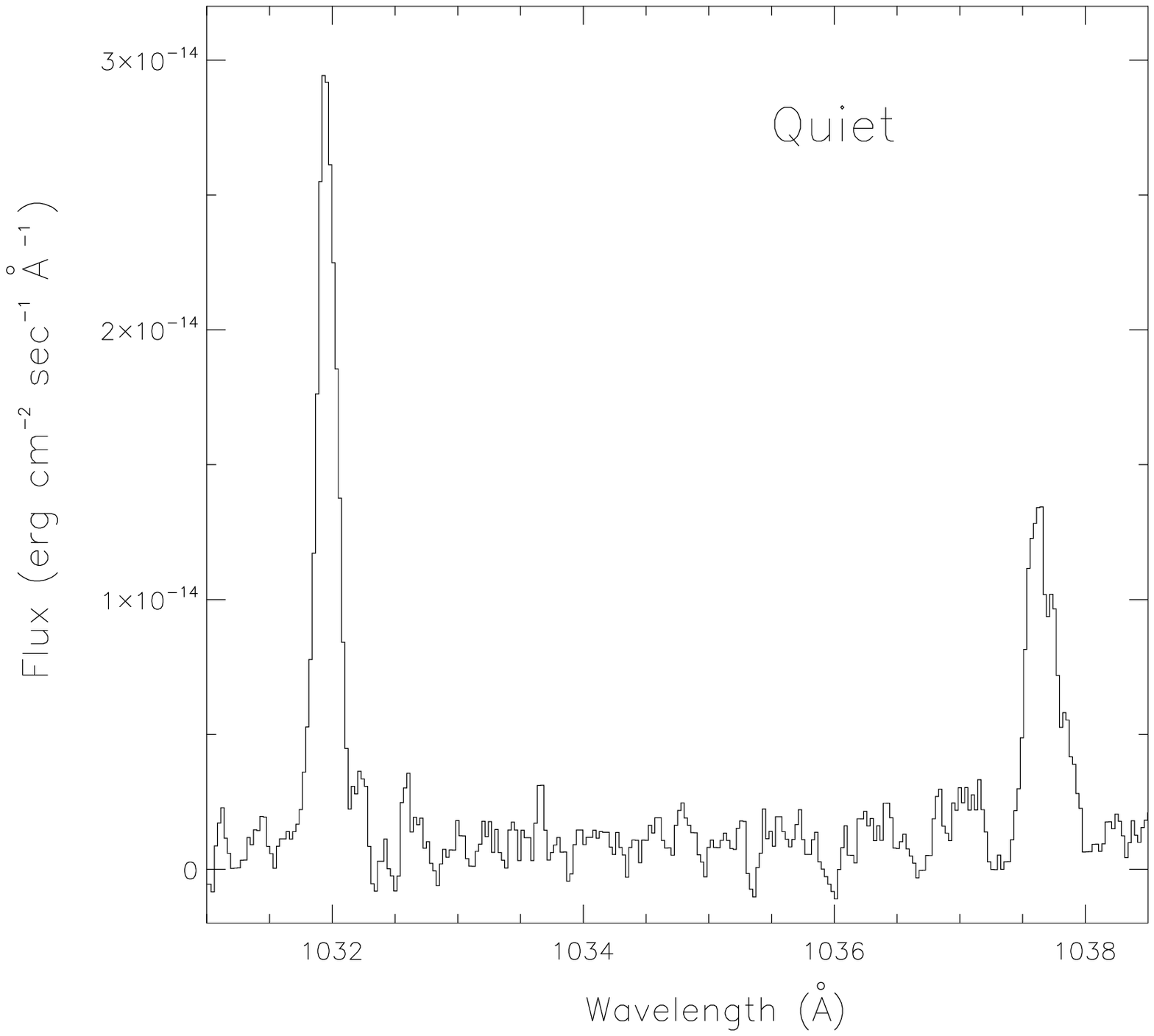}
\includegraphics[scale=0.45,angle=0]{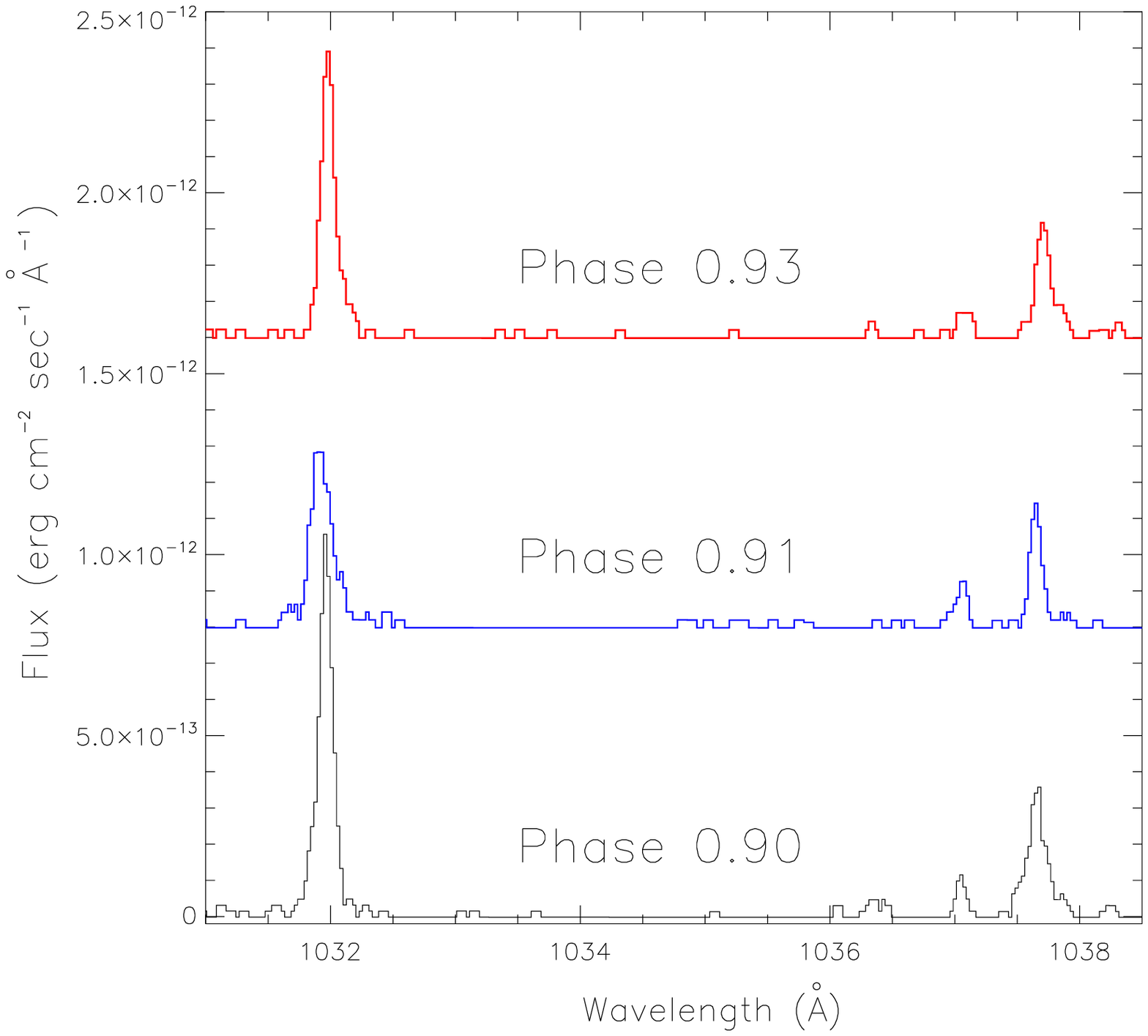}
\caption{Sample FUSE spectra of $\epsilon$ Eri containing the 1032 and 1038\,\AA\ transitions of \ion{O}{vi}. 
The left-hand panel shows the  quiet spectrum (1032/1038 intensity ratio = 1.93 $\pm$ 0.08), while the 
flare spectra at several orbital phases are given in the right-hand one. 
Phases are offset by 8$\times$10$^{-13}$  erg\,cm$^{-2}$\,s$^{-1}$ for ease of presentation. 
}
\end{figure*}

\begin{figure}
\includegraphics[scale=0.33,angle=90]{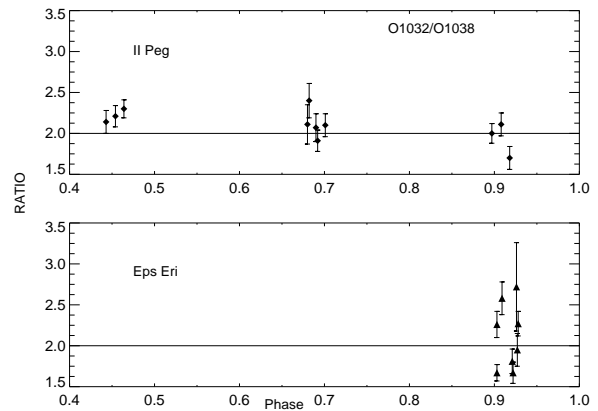}
\caption{Sample FUSE 1032/1038 line ratios in \ion{O}{vi} plotted as a function of orbital phase. Spectra are in time order as compared to the first part of the $\epsilon$\,Eri light curve and the total II\,Peg light curves.
}
\end{figure}

  \subsection{II Peg} 
   
 II\,Peg showed several spectra with measured \ion{O}{vi} 1032/1038 ratios greater than 2.0, but only  a few
 are statistically significant. For example, Specnum 04 has a measured ratio of 2.30 $\pm$ 0.11 (see Table 2),  larger than the theoretical value by 2.7$\sigma$, indicating a 99.3\%\ confidence level in our result being a true enhancement.  
The \ion{O}{vi} light curve is nearly constant over the 300 ksec time interval of the observations (except for a small brightening in the middle set of data; 135--155 ksec), and the count rate in the FUSE band indicates the star was fairly active during this period. 
 We computed the orbital phase of each spectrum using the ephemeris and rotational period of 6.72 days from 
Berdyugina et al. (1998). Our FUSE observations covered about 57\% of the orbital period, including phases
0.44--0.50, 0.68--0.70 and 0.88--0.90.  
 The 1032/1038 intensity ratio was enhanced for phases between 0.44--0.50, increasing from 2.14 to 2.30, and had a marginal enhancement at phase 0.91.  
We show sample spectra for the \ion{O}{vi} lines in Fig. 7,  and the 1032/1038 ratios are plotted as a function of orbital
phase in Fig. 6.

\begin{figure*}
\includegraphics[scale=0.45,angle=0]{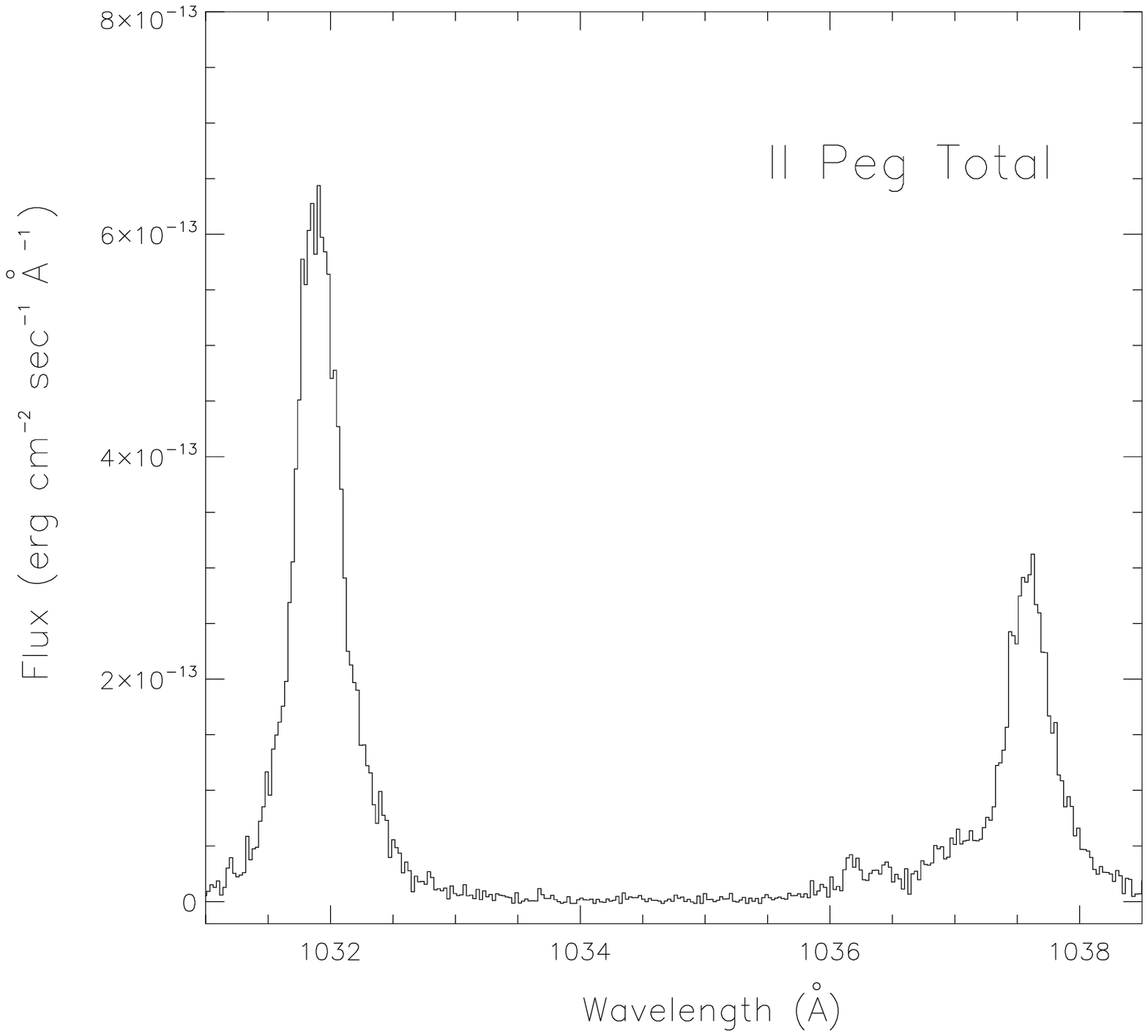}
\includegraphics[scale=0.45,angle=0]{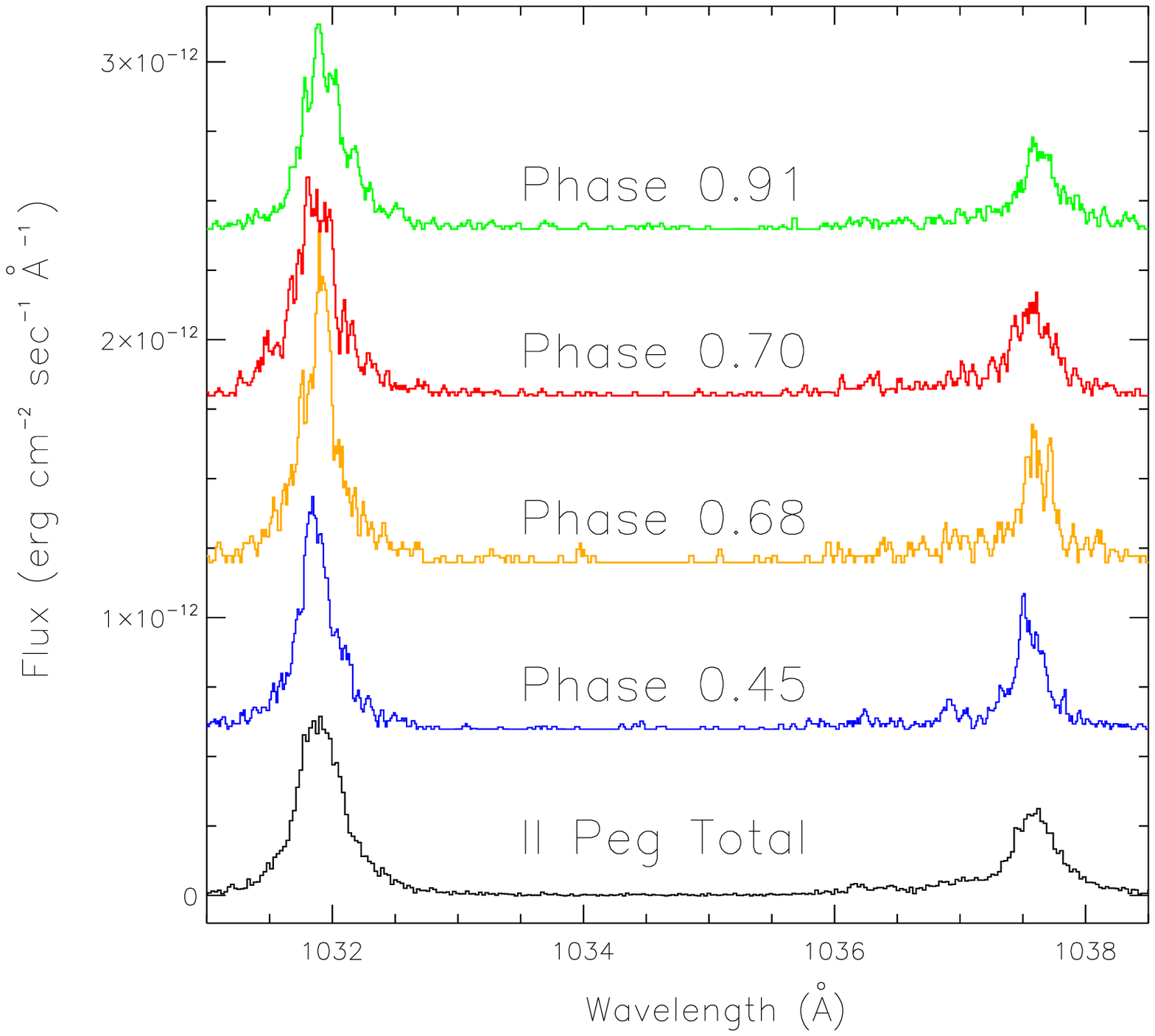}
\caption{Sample FUSE spectra of II\,Peg containing the 1032 and 1038\,\AA\ transitions of \ion{O}{vi}. 
The left-hand panel shows the total II\,Peg FUSE spectrum (1032/1038 intensity ratio = 2.03 $\pm$ 0.07), while selected orbital phases are given on the right.  Phases are offset by 6$\times$10$^{-13}$ erg\,cm$^{-2}$\,s$^{-1}$ for ease of presentation. 
}
\end{figure*}

\subsection{Prox Cen}

Prox Cen (GJ 551C) was observed by FUSE in 2003 with the LWRS aperture. 
Christian et al. (2004) found several 1032/1038 line ratios in the quiescent spectra which indicated the presence of significant opacity (i.e. values of $<$\,2.0), and used these to derive pathlengths for the TR of up to $\sim$\,13 km. 
However, these authors also measured 1032/1038 ratios in two flare spectra which were $>$\,2.0, even allowing for 
observational uncertainities.  These results are hence also included in Table 2.

\subsection{Densities and pathlengths}

To determine the column density of the \ion{O}{vi} region in our stellar sources, required for comparison with theory (see Section 3), we need estimates of both the density and pathlength. Unfortunately, there are few electron density ($n_e$) diagnostics in the FUSE spectral regions for ions formed at similar temperatures to \ion{O}{vi}, 
which has a temperature of maximum fractional abundance in ionisation equilibrium of T$_{e}$ = 10$^{5.5}$\,K (Bryans et al. 2009). The only possibility is \ion{Ne}{vi} (formed at T$_{e}$ = 10$^{5.6}$\,K; Bryans et al.), which has several density sensitive emission lines in the wavelength range $\sim$\,997--1010\,\AA. However, the features are weak and noisy in the FUSE datasets for $\epsilon$\,Eri and II\,Peg, and are not detected for Prox Cen. As a consequence, the \ion{Ne}{vi} lines provide at best only a constraint on the density range. For example, using the latest version (V6.0) of the 
{\sc chianti} database (Dere et al. 1997, 2009), the measured \ion{Ne}{vi} 
997.03/1005.7 ratio for II\,Peg of 1.1$\pm$0.6 indicates an electron density in the range $\sim$\,10$^{11}$--10$^{12.5}$\,cm$^{-3}$, while for $\epsilon$\,Eri the experimental 997.03/1005.7 ratio of 0.44$\pm$0.29 implies $n_e$ $\la$ 10$^{12}$\,cm$^{-3}$.  These are consistent with the earlier results of Byrne et al. (1987) and Jordan et al. (2001), who derived $n_e$ $\simeq$ 10$^{11.3}$ and 10$^{11.2}$\,cm$^{-3}$ for II\,Peg and $\epsilon$\,Eri, respectively, from a number of diagnostics. For Prox Cen, G\"{u}del et al. (2002) found a typical electron density of $\sim$\,10$^{11}$\,cm$^{-3}$ from \ion{O}{vii} lines. Given these results, we adopt $n_e$ = 10$^{11}$\,cm$^{-3}$ for all 3 stars, which should be reliable to within an order of magnitude.

Christian et al. (2006) have summarised estimates of the pathlengths ({\em l}) for the \ion{O}{vi} emitting regions in several late-type active stars, including Prox Cen. They find values ranging from around {\em l} = 10--10$^{2}$\,km, and we therefore adopt a typical pathlength of {\em l} = 30\,km, which should be accurate to an order of magnitude, similar to the electron density. Combining these, our expected column density $n_e${\em l} (in units of cm$^{-2}$) for the \ion{O}{vi} plasmas in our stellar sample should be around log $n_e${\em l} = 17.5.

\section{Discussion and future work}

First, we note that the simple detection of line intensity enhancements in several of our observations rules out a spherical geometry for the emitting plasmas, as theory predicts that a spherical region will show neither line intensity enhancement nor reduction, irrespective of the optical depth of the plasma (see Kerr et al. 2005 and Section 3). This result assumes that the observer is not close to the plasma surface, which is of course a valid assumption for distant astronomical sources.
Kerr et al. point out that their predictions for a spherical plasma may explain the few detections of opacity effects in stellar coronal emission lines, such as those of \ion{Fe}{xvii} (Matranga et al. 2005), \ion{O}{viii} and \ion{Ne}{x} (Testa et al. 2007). One might expect high temperature coronal emission to be distributed (approximately) uniformly over the stellar surface, particularly under quiescent conditions, hence leading to a spherical geometry. 

To further illustrate how the detection of intensity enhancement can constrain geometry we consider the theoretical case outlined in Section 3, i.e. that of an infinite plane slab with column density thickness $n_e${\em l}, observed at an angle $\theta$ to the perpendicular. Fig. 4 shows the 1032/1038 intensity ratio of \ion{O}{vi} plotted as a function of logarithmic column density for  different values of $\theta$. From the figure, we see that significant intensity enhancement is only predicted for a relatively narrow range of column density, log $n_e${\em l} $\simeq$ 16.5--18.0, compatible with the approximate value derived for the stellar sources in the present paper, log $n_e${\em l} $\simeq$ 17.5 (Section 4.4). Furthermore, the enhancement only occurs for small values of $\theta$, indicating that the \ion{O}{vi} emission regions are being observed close to face-on. In this regard, it is interesting to speculate that the increase in the 1032/1038 ratio with time measured during the 0.4--0.5 orbital period of II\,Peg (Fig. 6) may be due to the emitting plasma changing its orientation with respect to the line-of-sight such that $\theta$ decreases, resulting in an increase in the line intensity enhancement. However, alternatively the plasma parameters may be changing, hence leading to an increase or decrease in column density.

It can therefore be seen that the observation of line intensity enhancement in astronomical spectra due to opacity does allow some constraints to be placed on both the geometry of the emitting plasma and orientation to the observer.
However, further work is required. For example, it would be interesting to properly observe a stellar source at high spectral and temporal resolution over a full rotation period, to investigate if the line intensity enhancement can be correlated with rotation. This would provide evidence that the emitting region is changing its orientation with respect to the sightline. It would also be useful to extend the work to astronomical sources other than cool stars. 

Finally, although our research 
provides evidence for line intensity enhancements due to opacity in astrophysical sources,
it would be highly desirable to confirm the theory of Kerr et al. (2004, 2005) in a well-diagnosed laboratory plasma.
In particular, using a laboratory plasma of known geometry will allow a strigent test of the theory in terms of 
plasma geometry and orientation of the observer, as different results are expected depending on e.g. 
whether the plasma is slab, cylindrical or spherical (Kerr et al. 2004, 2005). 

Here we provide an outline design of an experiment using a high-power laser to check our theoretical modelling, through the observation of line ratios from a laboratory plasma that has been independently diagnosed. 
The experiment proposed below builds on recent work conducted using the NIKE laser (Back et al. 2006). In these studies, Back et al. demonstrated that it was possible to heat a low density (few mg\,cm$^{-3}$) aerogel foam target to electron temperatures of over 5$\times$10$^{7}$\,K with good uniformity. This was accomplished by driving a supersonic heating wave through the foam, producing a plasma of known density that disassembles through the progress of a rarefaction wave starting at the outside of the plasma. By doping the foam with an element in a central region of chosen geometry, a plasma of known, uniform temperature (diagnosed using Thomson scattering), uniform initial density, known composition and selected geometry can potentially be produced in the laboratory. The observation of a line intensity ratio (optically thick/optically thin) as a function of increasing dopant concentration  and angle will allow us to produce in the laboratory the analogue of the astrophysical situation described in Kerr et al. (2004, 2005), and thereby check our modelling. 

The schematic diagram of the experiment is shown in Fig. 8. 
A laser heats a block of aerogel foam at sub-critical density into which is doped the emitting material (for this excercise we choose the doped material to be sodium) in a region of a chosen (in this case planar) geometry. Thomson scattering determines the electron and ion temperature (both are needed: the first determines the kinetics and the second determines the line width and thereby the optical depth). The doped region remains unaffected until it is disrupted by the rarefaction wave that moves inwards from the outside of the foam. Calculations with the NIMP code of Rose (1997) indicate that for sodium doped into the central region, the kinetics reach a steady state in a few hundreds of picoseconds, which shows that the doped plasma has sufficient time to reach a steady-state before it is disrupted. The spectroscopic observation of the line intensity ratio is conducted at a variety of angles ($\theta$)
and for a number of different dopant concentrations. 

\begin{figure}
\includegraphics[scale=0.55,angle=0]{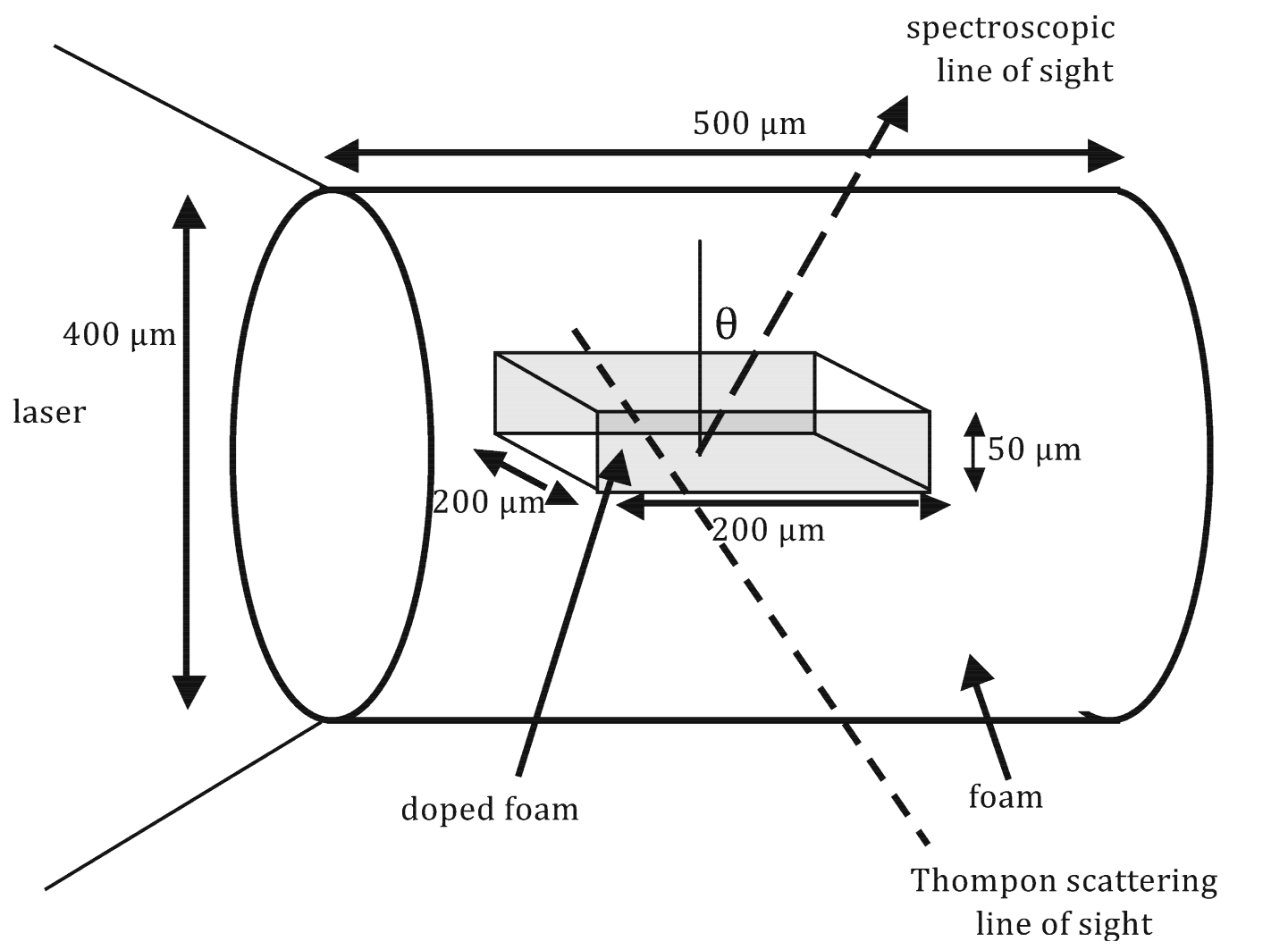}
\caption{
Schematic diagram of a possible experiment to measure line intensity enhancements arising from opacity effects in a laboratory plasma. See Section 5 for further details.
}
\end{figure}

\begin{acknowledgements}
 FPK is grateful to AWE Aldermaston for the award of a William Penney Fellowship.
DC thanks the Royal Society for an International Travel Grant in support of this research.
{\sc chianti} is a collaborative project 
involving researchers at the Naval Research Laboratory (USA), Rutherford
Appleton Laboratory (UK), and the Universities of Cambridge (UK), George Mason (USA) and Florence 
(Italy).
\end{acknowledgements}

\end{document}